# An Extension to DNA Based Fredkin Gate Circuits: Design of Reversible Sequential Circuits using Fredkin Gates

Himanshu Thapliyal and M.B Srinivas
(thapliyalhimanshu@yahoo.com, srinivas@iiit.net)
Center for VLSI and Embedded System Technologies
International Institute of Information Technology
Hyderabad-500019, India

## ABSTRACT

In recent years, reversible logic has emerged as a promising computing paradigm having its applications in low power computing, quantum computing, nanotechnology, optical computing and DNA computing. The classical set of gates such as AND, OR, and EXOR are not reversible. Recently, it has been shown how to encode information in DNA and use DNA amplification to implement Fredkin gates. Furthermore, in the past Fredkin gates have been constructed using DNA, whose outputs are used as inputs for other Fredkin gates. Thus, it can be concluded that arbitrary circuits of Fredkin gates can be constructed using DNA. This paper provides the initial threshold to building of more complex system having reversible sequential circuits and which can execute more complicated operations. The novelty of the paper is the reversible designs of sequential circuits using Fredkin gate. Since, Fredkin gate has already been realized using DNA, it is expected that this work will initiate the building of complex systems using DNA. The reversible circuits designed here are highly optimized in terms of number of gates and garbage outputs. The modularization approach that is synthesizing small circuits and thereafter using them to construct bigger circuits is used for designing the optimal reversible sequential circuits.

## 1.   INTRODUCTION

This section provides an effective background of reversible logic with its definition and the motivation behind it.

### 1.1 Definitions

Researchers like Landauer have shown that for irreversible logic computations, each bit of information lost generates $kT\ln 2$ joules of heat energy, where k is Boltzmann's constant and T the absolute temperature at which computation is performed [1]. Bennett showed that $kT\ln 2$ energy dissipation would not occur, if a computation is carried out in a reversible way [2], since the amount of energy dissipated in a system bears a direct relationship to the number of bits erased during computation. Reversible circuits are those circuits that do not lose information. Reversible computation in a system can be performed only when the system comprises of reversible gates. These circuits can generate unique output vector from each input vector, and vice versa, that is, there is a one-to-one mapping between input and output vectors. Thus, an NXN reversible gate can be represented as

$Iv=(I_1,I_2,I_3,I_4,\ldots\ldots\ldots\ldots\ldots I_N)$
$Ov=(O_1,O_2,O_3,\ldots\ldots\ldots\ldots\ldots O_N)$.

Where Iv and Ov represent the input and output vectors respectively. Classical logic gates are irreversible since input vector states cannot be uniquely reconstructed from the output vector states. There is a number of existing reversible gates such as Fredkin gate[3,4,5], Toffoli Gate (TG) [3, 4] and  TSG Gate [6,7].

### 1.2 Motivation

The reversible logic operations do not erase (lose) information and dissipate very less heat.  Recently, it was shown how to encode information in DNA and use DNA amplification to implement Fredkin gates [8]. Furthermore, in the past Fredkin gates were constructed using DNA, whose outputs are used as inputs for other Fredkin gates. Thus, it can be concluded that arbitrary circuits of Fredkin gates can be constructed using DNA. In order to design complex circuits using DNA, reversible designs of complex circuits should be available for Biochemists to work on. In the literature, researchers have so far addressed only the problem of designing reversible combinational circuits and no significant work has been addressed towards the design of reversible sequential circuits. Furthermore, reversible designs of sequential logic differ from combinational logic in that the output of the logic device is dependent not only on the present inputs to the device, but also on past inputs; *i.e.*, the output of a sequential logic device depends on its present internal state and the present inputs. This paper provides the initial threshold to building of more complex system having sequential circuits and which can execute more complicated operations using DNA. The novelty of the paper is the reversible designs of latches, Flip Flops, registers and other complex sequential circuits using Fredkin gate. Since, Fredkin gate has already been realized using DNA, it is expected that this work will initiate the building of complex systems using DNA.

## 2. FREDKIN GATE

Fredkin gate[3,4,5], a (3*3) conservative reversible gate as shown in Figure 1. It is called 3*3 gate because it has three inputs and three outputs. The term conservative means that the Hamming weight (number of logical ones) of its input equals the Hamming weight of its output. The input triple $(x_1, x_2, x_3)$ associates with its output triple $(y_1, y_2, y_3)$ as follows

$$y_1 = x_1$$
$$y_2 = (\neg x_1 \wedge x_2) \vee (x_1 \wedge x_3)$$
$$y_3 = (x_1 \wedge x_2) \vee (\neg x_1 \wedge x_3)$$

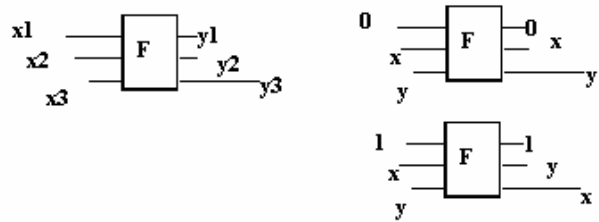

Figure 1. Fredkin Gate Symbol and Its working as a Conditional Switch

## 3. PROPOSED SEQUENTIAL CIRCUITS

The reversible D latch is built from the Fredkin gate, which is later used to design complex sequential circuits, as discussed in the section below.

### 3.1 Proposed Reversible D Latch Using Fredkin Gate
Figure 2 shows a conventional D latch. The characteristic equation of the D latch can be written as

$Q^+ = D \cdot E + \overline{E} \cdot Q$ . where D refers to the input and E can be considered as Enable or Clock Pulse

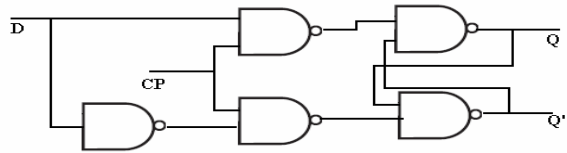

Figure 2. Conventional D Latch

The characteristic equation of the D latch can be mapped onto the Fredkin gate (F). Figure 3 shows the realization of the proposed reversible D latch. To avoid a fan-out problem, a Fredkin gate (F) is used to copy the output. It can be seen that the reversible D latch is highly optimized in terms of the number of reversible gates and garbage outputs (outputs that are not subsequently used).

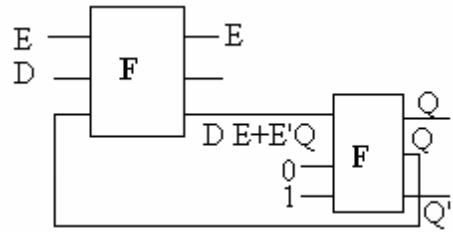

Figure 3. Reversible D latch built from Fredkin Gates

### 3.2 Complex Sequential Circuits Using D-Latch

The reversible D latch is used to implement more complex reversible sequential circuits. Figure 4 shows a reversible storage register constructed from four reversible D latches and a common clock input. Figure 5 shows the master-slave D flip flop designed from the proposed D Latch in which CP refers to the clock pulse.

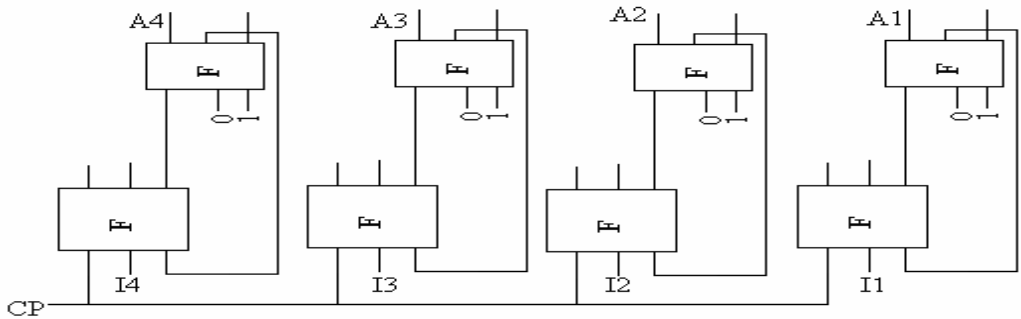

Figure 4. Reversible Register built from reversible D latches

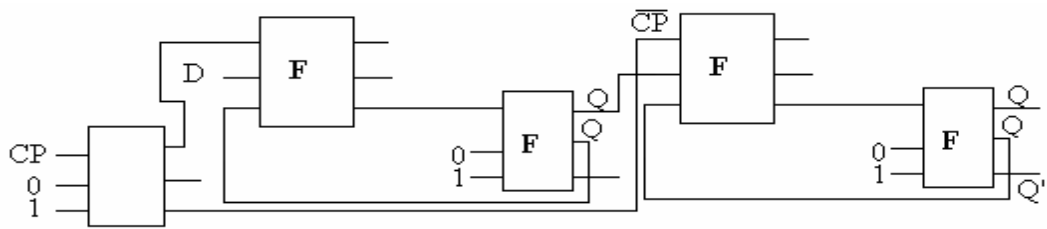

Figure 5. Proposed Reversible Master Slave D Flip Flop

Figure 6 shows a reversible shift register built from the proposed reversible master-slave D Flip Flop. Each clock pulse shifts the contents of the register one bit position to the right. All the proposed designs are highly optimized in terms of number of reversible gates and garbage outputs, the unused outputs in the figures are the garbage outputs.

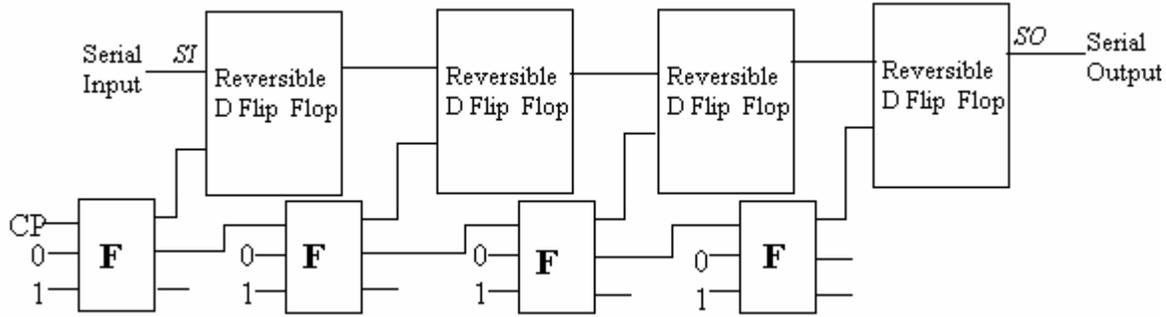

Figure 6. Reversible Shift Register built from reversible master-slave D flip flop

Figure 7 shows a reversible serial transfer of data from register A to register B using the proposed reversible master slave D flip flop. Figure 8 shows a reversible serial adder built from the proposed reversible sequential circuits. The two binary numbers to be added are stored in two shift registers. Bits are added one pair at a time, sequentially, through a reversible full adder. The other reversible combinational components required in the Figure 8 have already being proposed in the literature [9].

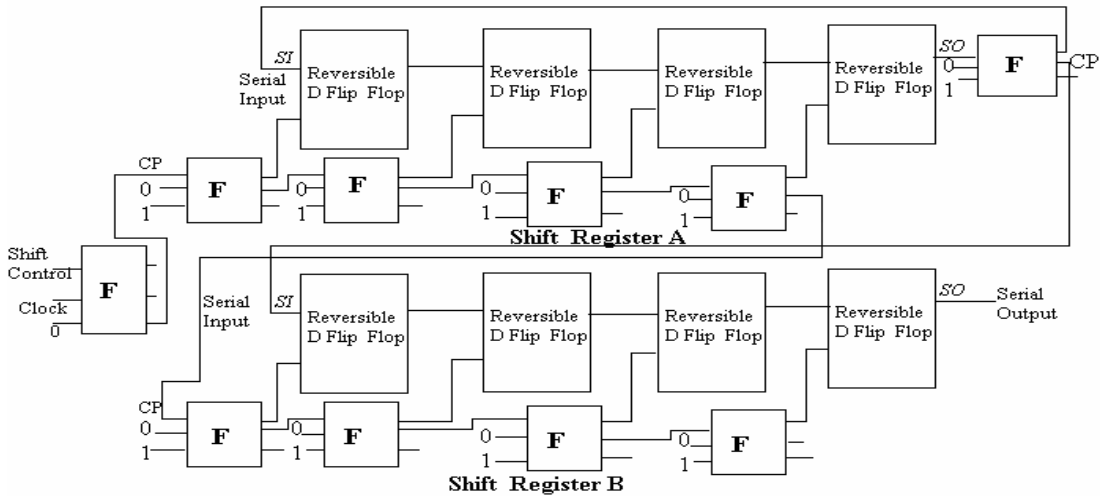

Figure 7. Serial Transfer from Register A to Register B using reversible master slave D flip flop

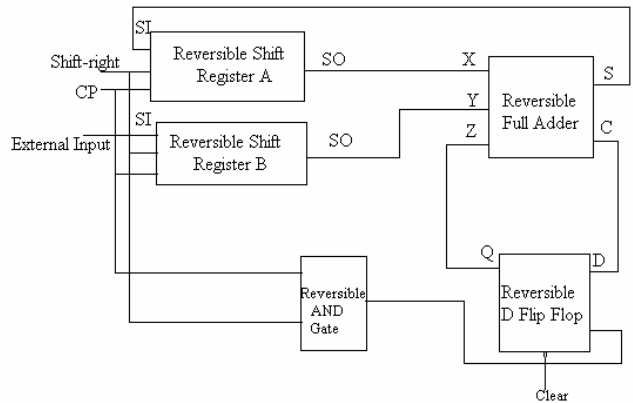

Figure 8. Serial Adder built From proposed Reversible D latch and Reversible Combinational Components

### 3.3 Derivation of Other Latches and Flip Flops Using Proposed D-Latch

The proposed reversible SR Latch is shown in Figure 9. It is derived from the design of the proposed reversible D-Latch. The block numbers in the Figure 9 shows the way the computation is proceeding to generate the output. Figure 10 shows the proposed reversible JK latch and Figure 11 shows the proposed reversible T-Latch derived from the proposed D-Latch. In the proposed designs, Fredkin gates are also used to copy the signals to avoid the fan-out problems.

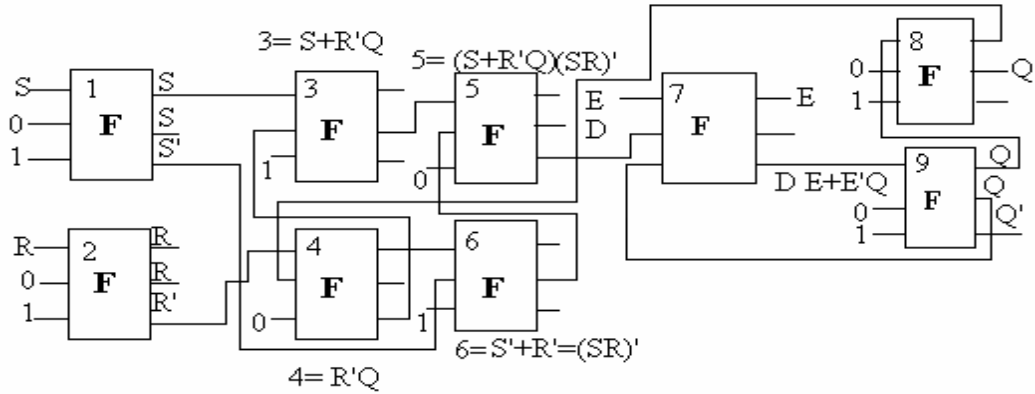

Figure 9. Proposed Reversible SR Latch

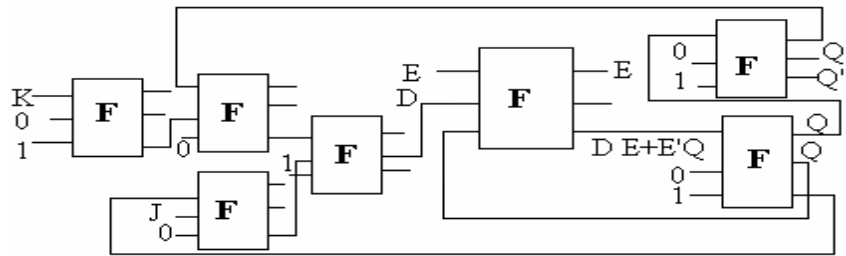

Figure 10. Proposed Reversible JK Latch

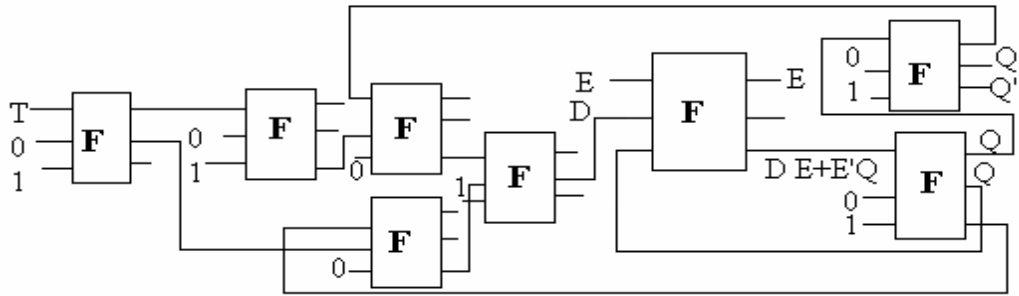

Figure 11. Proposed Reversible T Latch

The reversible Master Slave versions of the various Flip Flops are also designed using the proposed D-Latch. Figure 12 shows the reversible Master Slave SR Flip Flop designed using the proposed D-Latch, while Figure 13 and Figure 14 shows the proposed reversible Master Slave JK and Master Slave T Flip Flop respectively. Thus, the basic units required for the design of the complex sequential circuits have been provided, which can be appropriately used to design complex circuits.

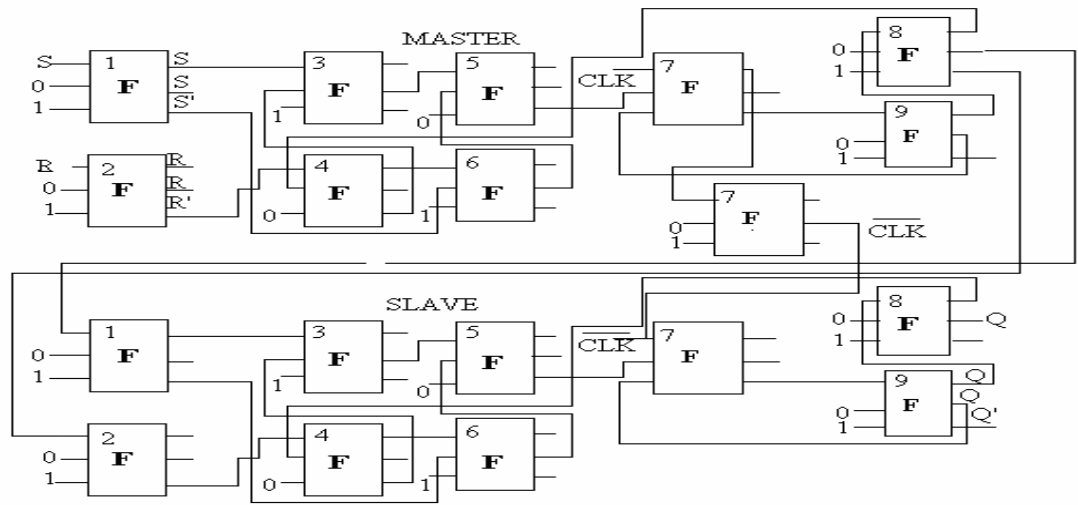

Figure 12. Proposed Reversible Master Slave SR Flip Flop

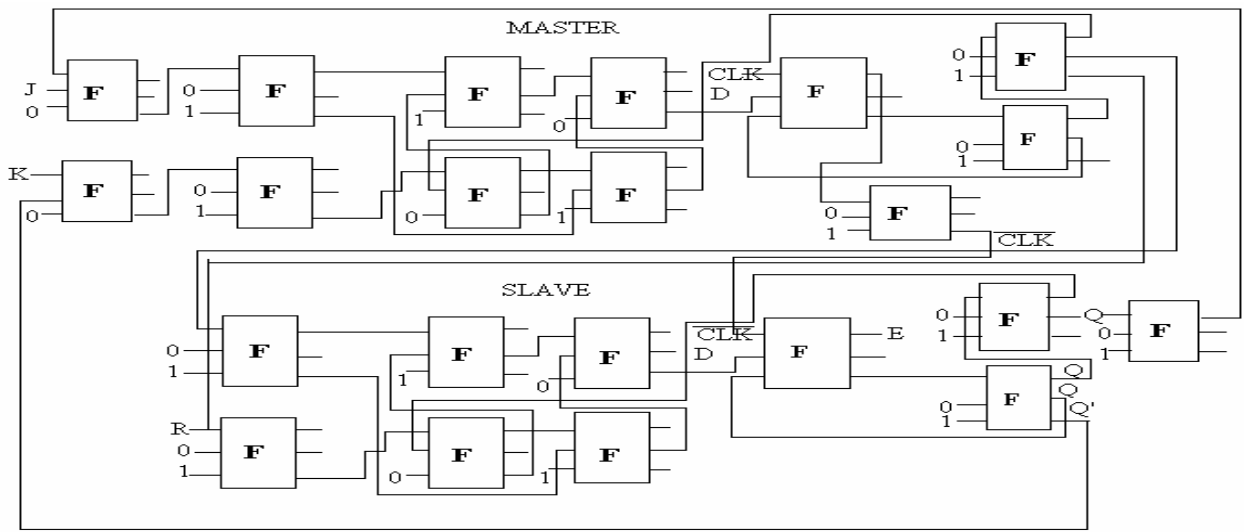

Figure 13. Proposed Reversible Master Slave JK Flip Flop

## 4. CONCLUSION

The focus of this paper is the design of the complex sequential circuits using Fredkin gates, to make it possible for the biologists and biochemists to design large reversible systems using DNAs. The reversible latches, Flip Flops, registers and other complex sequential circuits are designed using Fredkin gate. From a literature survey, we believe that this is the first work to propose a reversible latch and complex reversible sequential circuits. The design strategy is chosen in such a way to make them highly optimized in terms of number of reversible gates and garbage outputs.  This work forms an important step in the building of complex systems reversible systems using DNA computing.

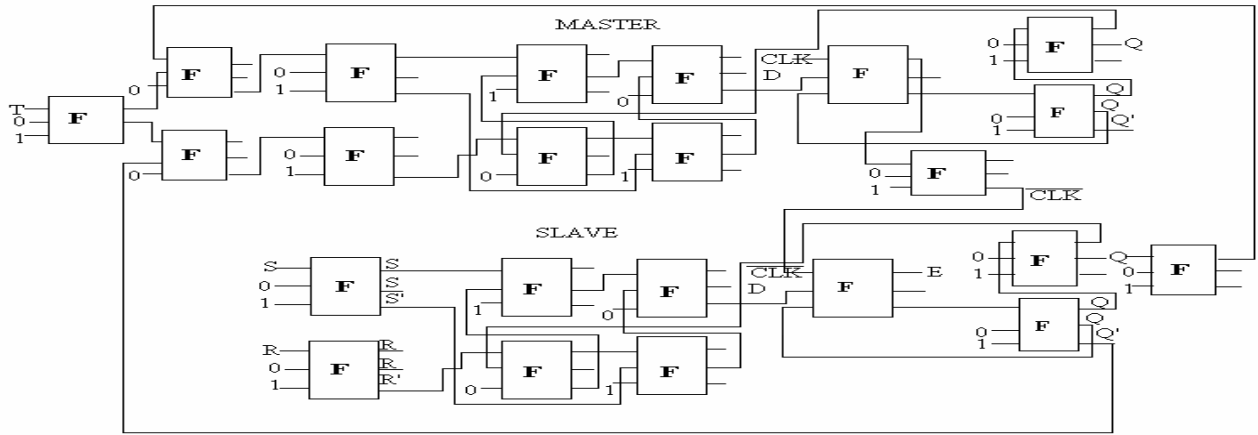

Figure 14.  Proposed Reversible Master Slave T flip Flop